\documentclass[
preprint, %For draft, 1-col layout, uncomment ``preprint" and ``onecolumn"
onecolumn,
nofootinbib,
superscriptaddress,
floatfix,
]{revtex4-1}
\usepackage{graphicx}
\usepackage{amsmath}
\usepackage{amssymb}

\usepackage{hyperref}
\hypersetup{
	colorlinks=true,
	linkcolor=blue,
	filecolor=magenta,      
	urlcolor=cyan,
}

\begin{document}
	
	\author{Thomas F. Varley$^*$}
	\affiliation{Vermont Complex Systems Center, University of Vermont, Burlington, VT, USA}
	\email{tfvarley@uvm.edu}
	
	\title{Considering dynamical synergy and integrated information; the unusual case of minimum mutual information.}

	\date{\today}
	
	\begin{abstract}
		This brief note considers the problem of estimating temporal synergy and integrated information in dyadic dynamical processes. One of the standard estimators of dynamic synergy is based on the minimal mutual information between sets of elements, however, despite it's increasingly widespread use, the mathematical features of this redundancy function have largely gone unexplored. Here, we show that it has two previously unrecognized limitations: it cannot disambiguate between truly integrated systems and disintegrated systems with first-order autocorrelation. Second, paradoxically, there are some systems that become \textit{more} synergistic when dis-integrated (as long as first-order autocorrelations are preserved). In these systems, integrated information can decrease while synergy simultaneously increases. We derive conditions under which this occurs and discuss the implications of these findings for past and future work in applied fields such as neuroscience. 
	\end{abstract}
		
	\maketitle 
	
	\section*{Motivation}
	
\begin{figure}
		\includegraphics[width=\textwidth]{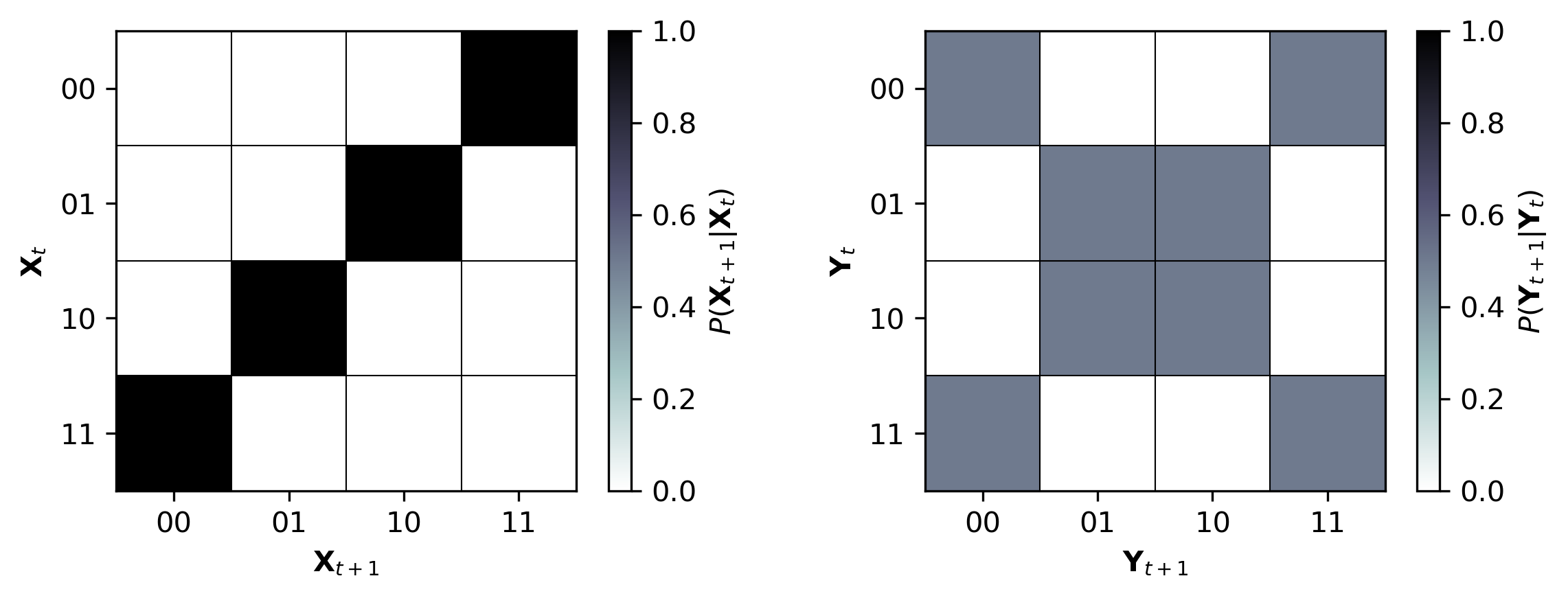}
		\caption{\textbf{Left:} The transition probability matrix for system \textbf{X}. This system is disintegrated: $X^1\bot X^2$, however, each individual process contains 1 bit of autocorrelation, and so the predictive power of the whole is trivially reducible to the sum of its parts. \textbf{Right:} The transition probability matrix for system \textbf{Y}. This system is ``integrated", in that there is 1 bit of temporal mutual information in the whole, but the individual elements contain no predictive power about their own futures.}
		\label{fig:tpms}
	\end{figure}

	Imagine two simple, boolean systems: $\textbf{X}=\{X^1,X^2\}$ and $\textbf{Y}=\{Y^1,Y^2\}$, with the following temporal structures. In \textbf{X}, the component processes are independent: $X^1\bot X^2$ and each $X^i$ oscillates $1\to0\to1\to\ldots$ with absolute determinism. \textbf{X} is not really a ``system" at all, but rather, two deterministic processes that happen to be next to each-other. In contrast \textbf{Y} is a truly synergistic system: as it evolves, the global parity $Y^1_t\bigoplus Y^2_t$ remains the same, but the specific micro-states consistent with the parity are chosen at random. For example, if $(Y^1_t, Y^2_t) = (0,0)$, then $(Y^1_{t+1}, Y^2_{t+1})$ could equal either $(0,0)$ or $(1,1)$ with equal probability, but never $(0,1)$ or $(1,0)$. Consequently, the individual $Y^i$ have no autocorrelation: the future $Y^i_{t+1}$ cannot be predicted from the state of $Y^i$, however, there is 1 bit of global autocorrelation: $I(\textbf{Y}_{t};\textbf{Y}_{t+1})=1$ bit. It is also worth nothing that \textbf{X} has the same macro-scale, parity-preserving property as well; if $\textbf{X}_t=(0,0)$, then $\textbf{X}_{t+1}=(1,1)$ both of which have even parity, and likewise if $\textbf{X}_t=(0,1)$ or $(1,0)$, the parity of the next state will odd forever. Consequently, we cannot say that any differences between the systems are attributable to some feature of the macro-state (parity) alone, but rather, some relationship between the macro-state and the micro-state. Transition probability matrices for both systems can be seen in Figure \ref{fig:tpms}.
	
	One simple and intuitive way we could try to differentiate \textbf{X} and \textbf{Y} is by asking ``how much greater is the whole than the sum of its parts?" More formally, how much more effectively can we predict the future of the whole system if we model it as a whole (i.e. with the true joint statistics) versus if we model it as a set of disconnected processes. This measure is typically referred to as $\Phi$, or $\Phi^{WMS}$ (where $WMS$ refers to ``whole-minus-sum") \cite{balduzzi_integrated_2008}:
	
	\begin{equation}
		\Phi^{WMS}(\textbf{X}) = I(\textbf{X}_t;\textbf{X}_{t+1}) - \sum_{i=1}^2 I(X^i_{t} ; X^i_{t+1})
	\end{equation}
	
	In the case of \textbf{X}, it is clear that $\Phi^{WMS}(\textbf{X})=0$ bit. The joint temporal mutual information is 2 bit, and the two marginal temporal mutual informations are each 1 bit, which is consistent with our intuition that \textbf{X} isn't really a system at all, but rather a kind of maximum entropy ``ideal gas." It contains information, but that information is not integrated in any meaningful way; the system is reducible to the sum of its parts. In contrast $\Phi^{WMS}(\textbf{Y})$ has a non-zero integrated information; 1 bit. This is because the joint state can be partially predicted (the parity will always the be the same, ruling out two possibilities), but the individual states are completely unpredictable. Not only is $\Phi^{WMS}(Y)>0$ bit, it is totally equal to $I(\textbf{Y}_t;\textbf{Y}_{t+1})$, indicating that \textit{all} of the temporal information is in the whole, and none is in the parts. So, while \textbf{X} is reducible, \textbf{Y} is not. This concept of integrated information has become a powerful measure of ``complexity" in complex systems \cite{mediano_integrated_2022}.
	
	\section*{Synergy, redundancy, and integrated information decomposition}
	
	These two toy models represent limiting extremes of information integration: total integration (\textbf{Y}) and total disintegration (also sometimes called segregation) (\textbf{X}). In the real world it is common to find systems that combine features of both extremes, combining integration and segregation in non-trivial ways; even systems that do integrate information typically aren't totally emergent the way \textbf{Y} is. It is almost always the case that $\Phi^{WMS}(\textbf{X})<I(\textbf{X}_t;\textbf{X}_{t+1})$, indicating that there is \textit{some} emergent, integrated information, but it doesn't characterize the entire structure (in fact, total integration seems to only be possible using logical XOR gates). Given this fact, it is natural to ask ``how much of the total information is integrated in the joint state of the whole, and how much is reducible to lower-order interactions?" That portion of the total information that is only present in the joint statistics and none of the lower-order ones is typically known as the \textit{synergy}, and the problem of parcellating out the total mutual information into different scales of interaction as the \textit{partial information decomposition} \cite{williams_nonnegative_2010}. 
	
	In the context of a multivariate dynamical process, we typically use a derivative of the partial information decomposition called the \textit{integrated information decomposition} ($\Phi$ID), which returns a taxonomy of all possible information dynamics in a given system \cite{mediano_towards_2021}. The logic of the $\Phi$ID is too complex to detail fully in this note, but there are several key intuitions that I want to highlight. The first is that the synergistic information in the context of the $\Phi$ID is all that information about the future joint state of $\textbf{X}_{t+1}$ that is only learnable when the joint state of $\textbf{X}_t$ is known. The persistent, collective nature of the temporal synergy has made it an area of interest for scientists interested in the philosophical idea of ``emergence", associating synergy with irreducible collective behaviors \cite{rosas_reconciling_2020,mediano_greater_2022}. The second key fact is that actually computing a numerical value for this number requires defining a \textit{redundancy function} that quantifies the information redundantly duplicated over all elements throughout past and future: intuitively, this would be the information that could be learned by observing $X^1_{t}$ or $X^2_{t}$ or  $X^1_{t+1}$, or $X^2_{t+1}$.
	
	There have been a handful of different redundancy functions that have been explored in the literature, however, the most widely applied to empirical datasets is the \textit{minimum mutual information} (MMI) function;
	
	\begin{equation}
		I_{\textnormal{Red}}(\textbf{X}_{t}) = \min_{i,j}I(X^{i}_t;X^{j}_{t+1})
	\end{equation}

	This function has a number of appealing properties. It is non-negative, has a closed-form estimator for normally distributed random variables, makes intuitive sense, and is based on pure information theory (not requiring any complex optimizations or concepts from other branches of mathematics). It also has a closed-form dual which directly quantifies the synergy without needing to bootstrap any intermediate values, which is almost unique among redundancy functions. 
	
	\begin{equation}
		I_{\textnormal{Syn}}(\textbf{X}) = (\textbf{X}_{t};\textbf{X}_{t+1}) - \max_{i}I(X^{i}_t;\textbf{X}_{t+1})
	\end{equation}
	
	This value also has the benefit of being non-negative and has a reasonably straightforward interpretation: it is the information about the future of the whole that is not disclosed by the most informative part. In the scientific literature, the MMI-based integrated information decomposition is far and away the most popular, having been applied to evolutionary analysis of boolean networks \cite{varley_evolving_2024}, \textit{in-silico} models of spiking neural networks \cite{menesse_integrated_2024}, macaque \cite{gatica_transcranial_2024} and human \cite{luppi_synergistic_2022} brain data, analysis of physical phase transitions \cite{mediano_integrated_2022}, and clinical studies of loss of consciousness in anesthesia or brain injury \cite{luppi_reduced_2023,luppi_synergistic_2024}. Despite its increasingly widespread use, the MMI-based redundancy framework has gone relatively unstudied from a mathematical perspective. In the multivariate information theory literature, there is a long history of analyzing different operational definitions of redundancy in the context of the single-target mutual information, but comparatively less work done on multi-target redundancy. 
	
	\section*{MMI-based measures cannot disambiguate between different systems}	
	
	How would we expect these measures of redundancy and synergy to behave in our two extreme systems? Both systems, \textbf{X} and \textbf{Y}, have 0 bit of redundancy. This makes sense: in \textbf{X} there is no integration at all, since $X^1\bot X^2$, while in \textbf{Y}, the individual futures are randomized and the macro-scale parity is only visible at the level of the ``whole." Where the situation breaks down is when we consider synergy. Intuitively, we would expect \textbf{X} to have no synergy (since the whole is totally reducible to the sum of its parts, as demonstrated by the $\Phi^{WMS}$ measure), and for \textbf{Y} to have maximal synergy. This is true in the case of \textbf{Y}; $I_{\textnormal{Syn}(\textbf{Y})} = 1$ bit, accounting for 100\% of the total temporal mutual information. Unfortunately, $I_{\textnormal{Syn}(\textbf{X})} = 1$ bit as well, suggesting that a disintegrated system somehow has 50\% of its total temporal mutual information in the ``irreducible" synergy atom. 
	\newline \newline 
	What is happening here? 
	\newline \newline 
	We have already seen that $\Phi^{WMS}(\textbf{X})=0$ bit, meaning that the joint mutual information neatly factors:
	
	\begin{equation}
		I(\textbf{X}_t;\textbf{X}_{t+1}) = I(X^1_{t};X^1_{t+1}) + I(X^2_{t};X^2_{t+1})
	\end{equation}

	By the chain rule of mutual information, we can re-write this as:
	
	\begin{align}
		I(\textbf{X}_t;\textbf{X}_{t+1}) &= I(X^1_t;\textbf{X}_{t+1}) + I(X^2_t;\textbf{X}_{t+1} | X^1_t).
	\end{align}

	Since we know that each $X^1\bot X^2$, then the conditional mutual information reduces; $I(X^2_t;\textbf{X}_{t+1} | X^1_t) = I(X^2_t;\textbf{X}_{t+1})$ and so:
	
	\begin{equation}
		\label{eq:factored}
		I(\textbf{X}_t;\textbf{X}_{t+1}) = I(X^1_{t};\textbf{X}_{t+1}) + I(X^2_{t};\textbf{X}_{t+1})
	\end{equation}

	The structure of the $I_{\textnormal{Syn}}$ function means that, even though the two $X^i$ are independent, the smaller of the two mutual information terms in Equation \ref{eq:factored} will be counted as synergy. This information clearly isn't synergistic in the way that we understand it, however. In a sense it is synergy: it's that information about the future of $\textbf{X}_{t+1}$ that can only be learned when the states of both $X^i$ are known, but it doesn't necessarily reflect any non-trivial "integration" (which, from the $\Phi^{WMS}$ results, we know doesn't actually exist!).
	\newline \newline 
	\textit{If one were presented with an unknown system and told that it had 0 bit of redundancy and 1 bit of synergy, one would have no way to knowing if the mystery system actually was truly synergistic, or if it was a case of spurious synergy, as both \textbf{X} and \textbf{Y} have the same ``profile" according to the MMI redundancy.}
	
	\section*{Paradoxical increases in synergy}
	
	Even more alarmingly, for a given pair $X^1$ and $X^2$ with fixed autocorrelations, the amount of synergy can actually increase when $X^1$ and $X^2$ are independent. The logic is straightforward. To keep things simple, let's specify that $X^1$ is the variable that satisfies $\max_i I(X^i;\textbf{X}_{t+1})$. Then:
	
	\begin{align}
		I_{\textnormal{Syn}}(\textbf{X}) &= I(\textbf{X}_t ; \textbf{X}_{t+1}) - I(X^1_t;\textbf{X}_{t+1}) \\
		%&= I(X^1_{t};\textbf{X}_{t+1}) + I(X^2_{t};\textbf{X}_{t+1}|X^1_{t}) - I(X^1_{t};\textbf{X}_{t+1}) \\ 
		%&= I(X^2_{t};\textbf{X}_{t+1}|X^1_{t}) \\ 
		%&= H(X^2_t | X^1_t) + H(\textbf{X}_{t+1}| X^1_t) - H(X^2_t, \textbf{X}_{t+1}|X^1_{t}) \\
		%&= H(X^2_t | X^1_t) + H(\textbf{X}_{t+1}| X^1_t) - H(X^2_t|X^1_t) - H(\textbf{X}_{t+1}|X^1_t, X^2_t) \\
		&= H(\textbf{X}_{t+1}| X^1_t) - H(\textbf{X}_{t+1}|X^1_t, X^2_t) \label{eq:syn_ent} \\
		&= I(X^2_{t} ; \textbf{X}_{t+1} | X^1_t) \label{eq:syn_mi}
	\end{align}

	Equations \ref{eq:syn_ent} and \ref{eq:syn_mi} give us the interpretation of the minimum mutual information-based synergy in terms of entropy: it is the difference between our uncertainty about the joint future after learning just  $X^1_t$ (the maximally informative part) and the uncertainty about the joint future after learning both $X^1_t$ and $X^2_t$ simultaneously. How does this quantity change if $X^1$ and $X^2$ are independent versus if they are coupled? 
	
	If we define $\textbf{X}^{\bot}$ as the twin of system \textbf{X}, but with $X^1\bot X^2$. In $\textbf{X}^{\bot}$, each $X^i$ has the same first-order autocorrelations as in the original \textbf{X}, but no multi-element integration. $\Phi^{WMS}(\textbf{X}^{\bot})=0$ bit. To more directly assess the effects of disintegration, we can compute:
	
	\begin{equation}
		\Delta I_{\textnormal{Syn}}(\textbf{X}^{\bot}, \textbf{X}) = I_{\textnormal{Syn}}(\textbf{X}^{\bot}) - I_{\textnormal{Syn}}(\textbf{X})
	\end{equation}

	This measure will tell us how much synergy increases or decreases when \textbf{X} is disintegrated into $\textbf{X}^{\bot}$. If $\Delta I_{\textnormal{Syn}}(\textbf{X}) \geq 0$, then the minimum mutual information measure is, in some sense, fatally compromised as it would show a case of a disintegrated system with $\Phi^{WMS}=0$ bit, that nevertheless appeared to have non-trivial amounts of emergent, synergistic information. After forcing independence in $\textbf{X}^{\bot}$, we can directly write out it's $I_{\textnormal{Syn}}$ function:
	
	\begin{align}
		I_{\textnormal{Syn}}(\textbf{X}^{\bot})  &= H(X^2_{t+1}) - H(X^2_{t+1}| X^2_t) \\
		&= I(X^2_t ; X^2_{t+1})
	\end{align}

	If we subtract $I_{\textnormal{Syn}}(\textbf{X}^{\bot}) - I_{\textnormal{Syn}}(\textbf{X})$: 
	
	\begin{align}
		\Delta I_{\textnormal{Syn}}(\textbf{X}^\bot, \textbf{X}) %&= I(X^2_t ; X^2_{t+1}) - I(X^2_{t} ; \textbf{X}_{t+1} | X^1_t) \label{eq:delta_1} \\
		%&= H(X^2_t) - H(X^2_t|X^2_{t+1}) - H(X^2_t|X^1_t) + H(X^2_t|X^1_{t+1},X^2_{t+1},X^1_t) \\
		&= I(X^1_t;X^2_t) - I(X^1_t;\textbf{X}_{t+1}|X^2_t) \label{eq:delta_2}
	\end{align}

	It is not the case that $\Delta I_{\textnormal{Syn}}(\textbf{X})$ is strictly non-negative, nor is it strictly non-positive. This means that there are cases when disintegrating the system increases the apparent synergy and cases where it decreases it. Equation \ref{eq:delta_2} can give us insight into when this happens; it describes the change in synergy as a function of the \textit{instantaneous pairwise} correlation. If $X^1$ and $X^2$ are strongly functionally connected, that can ``swamp" the information that $X^1_t$ discloses about the joint future, conditional on $X^2_t$ and the MMI-synergy will be increased under a null model that preserves the autocorrelation. In contrast, systems that are weakly correlated will see the MMI-synergy destroyed by the same intervention. If we recall system \textbf{Y}, the synergistic, parity-preserving system, we can see this in action. Both $Y^1$ and $Y^2$ have zero autocorrelation, and no functional connectivity. Consequently, when the individual timeseries are permuted the synergy (and, in fact, all the temporal mutual information) falls to 0. 
	
	\section*{Application to brain data}
	
	\begin{figure}
		\includegraphics[width=\textwidth]{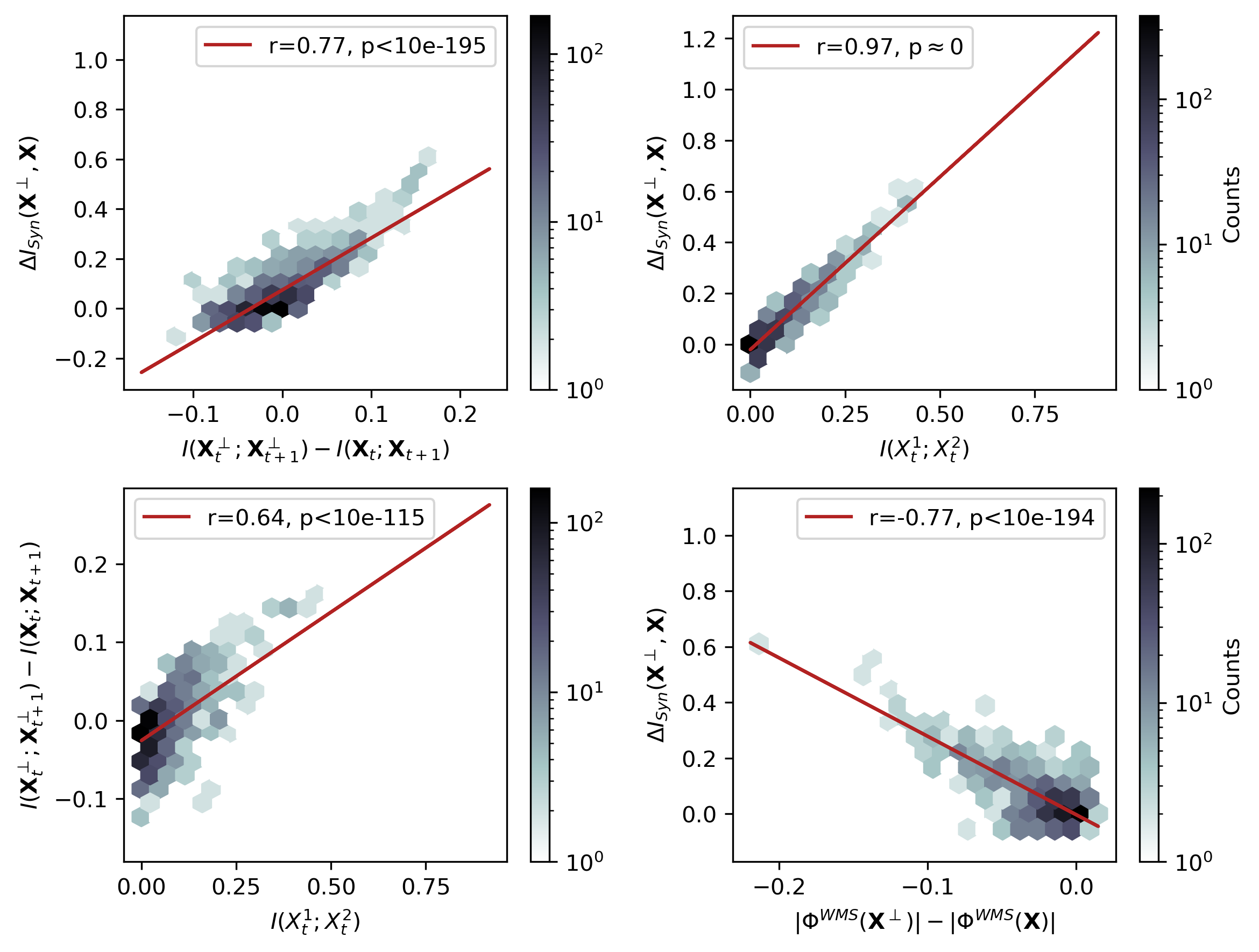}
		\caption{\textbf{Top left:} We find a strong, positive relationship between the change in total temporal mutual information upon disintegration and the change in synergy. Systems that show an increase in temporal mutual information also tend to show an increase in synergy. \textbf{Top right:} As expected given the derivation of Equation \ref{eq:delta_2}, the correlation between $X^1$ and $X^2$ is a significant driver how the apparent synergy in $I(\textbf{X}_t;\textbf{X}_{t+1})$ changes after disintegration. The greater the functional connectivity, the more synergy is produced when doing the circular shifts. \textbf{Bottom left:} The relationship between functional connectivity and global increase in temporal mutual information upon disintegration is strong, but not as strong as the change in synergy. \textbf{Bottom right:} This plot illustrates the central paradox discussed here: after disintegration, the integrated information $\Phi^{WMS}(\textbf{X})$ goes to zero, however in many pairs of brain regions, the MMI-synergy increases in a predictable, approximately linear way. Clearly synergy and integrated information are behaving in profoundly different ways.}
		\label{fig:hexbins}
		\end{figure}
	
	The MMI-synergy has been most prominently explored in the context of brain data - specifically functional neuroimaging. To better get a sense of how the properties of MMI detailed here might contextualize these results, here we will briefly explore how they manifest in empirical data. We will use one subject from the Human Connectome Project \cite{van_essen_wu-minn_2013} (these data were previously published in \cite{varley_partial_2023,varley_multivariate_2023}). We will sample 1,000 unique pairs of brain regions, and compare $I_{\textnormal{Syn}}\{(X^1,X^2\})$, $I_{\textnormal{Syn}}(\{X^1,X^2\}^{\bot})$, $\Delta I_{\textnormal{Syn}}(\{X^1,X^2\},\{X^1,X^2\}^{\bot})$, and functional connectivity. The disintegrated, autocorrelation-preserving nulls were done using a circular-shift permutation, averaged over 100 permutations. All results can be visualized in Figure \ref{fig:hexbins}.
	
	The first striking result is that, in about $37.5\%$ of sampled pairs, the autocorrelation-preserving null actually increased the total temporal mutual information. If we consider temporal mutual information to be a measure of how much ``structure" a dynamic system has, we are faced with the unusual circumstance where decreasing integration increased the apparent structure. Mathematically this is not necessarily problematic (it can occur when the entropy of the immediate future $H(\textbf{X}_{t+1})$ increases more than the entropy of the future conditioned on the past $H(\textbf{X}_{t+1}|\textbf{X}_t)$, which can occur when the $X^i$ are autocorrelated), but it does show that care must be taken when considering the relationship between first-order and higher-order autocorrelations. 
	
	If we correlate the change in synergy $\Delta I_{\textnormal{Syn}}(\textbf{X}, \textbf{X}^{\bot})$ with the change in temporal mutual information, we find a strong and highly significant positive correlation: $r=0.77$, $p<10^{-195}$. This means that, in systems where disintegration increases the total temporal mutual information, the ``added" information is taken to be synergistic. Despite this injection of synergy, disintegrating the systems almost always caused $\Phi^{WMS}(\textbf{X})$ to go to zero, suggesting a profound disconnect between ``information integration" as interpreted by $\Phi$ and ``information integration" as interpreted by MMI-synergy. This is something of a problem from the MMI-based approach, as the whole architecture of the $\Phi$ID was introduced to be a refinement of the original approach to integrated information theory.
	
	Finally, we experimentally confirm the relationship revealed in Equation \ref{eq:delta_2}: the change in synergy induced by autocorrelation-preserving disintegration is overwhelmingly positively correlated with the pairwise mutual information between elements ($r=0.97$, $p\approx 0$). While a similar positive correlation was observed between pairwise functional connectivity and the increase in temporal mutual information, it was only about half a strong ($r=0.64$, $p<10^{-115}$), confirming that the link between pairwise functional connectivity and synergy is analytically direct and not simply mediated by the overall change in temporal mutual information. 
	
	\section*{Where to go from here}
	
	Speaking just as a single author, I feel strongly that, despite its convenience and ease of computation, that the minimum mutual information function should probably not continue to be used in its current form. The inability to disambiguate between qualitatively different extremes of integration and disintegration, as well as the paradox that disintegrating a highly-autocorrelated system can actually increase the apparent synergy makes its interpretation difficult when considering empirical data where the ground truth is unknown. 
	
	Other redundancy functions may solve these problems. For example, an alternative might be the $I_{\textnormal{Red}}^{\tau sx}$ function \cite{varley_decomposing_2023}, which takes an alternative perspective on redundancy; one based on local information theory and overlapping exclusions of probability mass. When it was introduced, $I_{\textnormal{Red}}^{\tau sx}$ was applied to the disintegrated and integrated toy systems described above and was able to distinguish between them, unlike the MMI. Like any measure, it has its pros and cons: in addition to disambiguating between the different systems, it is also differentiable. It can, however result in negative redundancy and synergy atoms, which can be difficult to interpret. While originally developed strictly for discrete probability distributions, an elegant expansion to differential mutual information was recently introduced \cite{ehrlich_partial_2024}, which could make the redundancy function more universally accessible. Alternately, one might consider attempting to generalize the specificity and ambiguity decompositions from Finn and Lizier \cite{finn_pointwise_2018} to the problem of multiple targets. Finally, previous work on $\Phi$ID has made use of the common change in surprisals (CCS) \cite{mediano_towards_2021}, although this function has been critiqued as an approach to redundancy \cite{varley_partial_2023}.
	
	Alternatively, other frameworks beyond the $\Phi$ID could be explored. All approaches based on the partial information decomposition (including the $\Phi$ID described here, as well as derivatives like the partial entropy decomposition \cite{ince_partial_2017} and the generalized information decomposition \cite{varley_generalized_2024}) suffer from being ``redundancy-first" approaches, and so are rather oblique ways to assess the synergy. Instead, synergy-first approaches, such as the synergistic disclosure framework \cite{rosas_operational_2020} or the recently-introduced $\alpha$-synergy decomposition \cite{varley_scalable_2024} may provide a more direct route to understanding synergistic information.
	
	Finally, this is just the latest in a recent string of results suggesting that the relationship between synergy and maximum entropy ``randomness" is deeper than it appears at first blush. Generally, maximizing entropy is associated with loss of information, however in the case of higher-order synergies, that relationship is more nuanced. The first major finding, from Orio et al., showed that adding stochastic noise to the dynamic function of an elementary cellular automata can increase the synergy \cite{orio_dynamical_2023}. Subsequently, Varley and Bongard showed that random Boolean networks are inherently synergistic, and that there were deep dynamical similarities between random networks and synergistic ones that did not apply to redundant systems \cite{varley_evolving_2024}. Now we find that disintegrating multivariate dynamical systems can also increase the synergy (for a given definition of synergy). The relationship between synergistic information and unstructured entropy may challenge our understanding of how ``information" and ``uncertainty" are related, and partly explain why the problem of non-negative information decomposition has proven so stubbornly difficult. 

	\section*{Conclusions}
	
	In this short note, we have explored the nature of the minimum mutual information temporal redundancy function and its relationship with dynamic synergy. We found that MMI-based analysis cannot disambiguate between two very different systems; one with ``true" synergy, and a disintegrated system with zero integrated information but non-trivial first-order autocorrelations. Both systems have zero redundancy, and one bit of synergy, even though only one system has non-zero integrated information.
	
	We then showed that the MMI-based synergy has a peculiar property: in some cases, disintegrating a system (forcing the two halves to be independent) while preserving lower-order autocorrelation can actually increase the apparent synergy. Destroying integrated information can inject synergy; a paradox. This presents something of a problem for future work on temporal synergy and redundancy: how can we disambiguate between that synergy that truly reflects a ``greater than the sum of its parts" integration of information, versus that synergy that is being driven by independent processes with high autocorrelation? Understanding these results in deeper mathematical detail may help elucidate the curious, emerging relationship between synergy and entropic randomness, and in doing so, broaden our understanding of emergent information in multivariate systems. 
	
	%\bibliography{mmi_critique}

\end{document}